\documentclass[a4paper,fleqn,usenatbib]{mnras}
\usepackage{amsmath}
\usepackage{amssymb}
\usepackage{xspace}
\usepackage[normalem]{ulem}
\usepackage{pifont}

\usepackage{natbib}
\usepackage{graphicx}

\def\yr{{\rm yr}} 
\def\hr{{\rm hr}}

\def\m{{\rm m}} 
\def\cm{{\rm c}\m} 

\def\Jy{{\rm Jy}} 

\def\mas{{\rm mas}} 

\title[Differentiating Wind and Black Hole Driven Jets]{
  Differentiating Disk and Black Hole Driven Jets
  with EHT Images of Variability in M87
}

\author[Jeter et al.]{
  Britton~Jeter,$^{1,2,4}$
  Avery~E.~Broderick,$^{1,2,4}$
  and Roman~Gold$^{3,4}$
\\
$^{1}$Department of Physics and Astronomy, University of Waterloo, 200 University Avenue West, Waterloo, ON N2L 3G1, Canada\\
$^{2}$Waterloo Centre for Astrophysics, University of Waterloo, Waterloo, ON N2L 3G1, Canada\\
$^{3}$Institut f{\"u}r Theoretische Physik, Johann Wolfgang Goethe-Universit\"at, Max-von-Laue-Stra{\ss}e 1, 60438 Frankfurt, Germany\\
$^{4}$Perimeter Institute for Theoretical Physics, 31 Caroline Street North, Waterloo, ON N2L 2Y5, Canada}

\date{Accepted XXX. Received YYY; in original form ZZZ}

\begin{document}
\label{firstpage}
\pagerange{\pageref{firstpage}--\pageref{lastpage}}
\maketitle

\begin{abstract}

	Millimetre-wavelength very long baseline interferometric (mm-VLBI) observations of M87 by the Event Horizon Telescope (EHT) should provide a unique opportunity to observe and characterize the origins of jet variability already seen at longer wavelengths.  Synchrotron spot models have been used to model variability near black holes; this work extends these by allowing spots to shear and deform in the jet velocity field.  Depending on the position of the spot, shearing forces can significantly alter the structure of the spot, producing distinct signals in reconstructed images and light curves.  The maximum intensity of the shearing spot can vary by as much as a factor of five depending on the spot azimuthal launch position, but the intensity decay time depends most significantly on the spot radial launch position.  Spots launched by a black hole driven jet exhibit distinct arc structures in reconstructed images, and exhibit brighter and shorter-lived enhancements of the light curve.  Spots launched by a wind-driven jet have exhibit much simpler structures in the image, and longer-lived light curve enhancements than spots launched by a black hole driven jet.   

\end{abstract}


\section{Introduction}

Of the most luminous extragalactic radio sources, nearly every one shares a similar double-lobe radio structure, with a bright radio galaxy or quasar sitting between the two lobes.  These active galactic nuclei (AGN) all exhibit narrow, highly beamed relativistic jets with typical Lorentz factors of 10 to 100 and extend over wide range of intergalactic distances, from a few kpc to tens of Mpc.  Assuming the large scale radio lobes are powered by the termination shock of these jets against the intergalactic medium, it is possible to estimate their energetic content to lie between $10^{42}~{\rm erg\,s^{-1}}$ and $10^{48}~{\rm erg\,s^{-1}}$ \citep{LongHEA:11}.  At these values, the power output of the central AGNs in the form of relativistic jets rivals their electromagnetic luminosities.
This extraordinary kinetic luminosity plays a central role in the evolution of galaxies and galaxy clusters.  Via radio-mode feedback, AGN jets inject both energy and momentum in the circumgalactic and intracluster medium, suppressing star formation \citep{ McNam-Nul:12, Fab:12}.  Thus, the origin and content of AGN jets remain a critical input to understanding the evolution of structure in the universe.  This, in turn, requires a well-developed model for how relativistic jets are launched and how this depends on the properties of the parent AGNs, e.g., the evolution of black hole spin \citep{Worrall:09}.

The Event Horizon Telescope (EHT) is a global millimetre-wavelength, very-long baseline interferometer (mm-VLBI), capable of generating the first images of an AGN and resolving the horizons of the central supermassive black holes \citep{EHT-1:19, EHT-2:19, EHT-3:19, EHT-4:19, EHT-5:19, EHT-6:19}.  Currently, the EHT is comprised of seven telescopes located at five sites: the Submillimeter Array (SMA) and James Clerk Maxwell Telescope (JCMT) in Hawaii, the Arizona Radio Observatory Submillimeter Telescope (ARO-SMT) on Mt.~Graham, the Large Millimeter Telescope (LMT) in Mexico, the Atacama Large Millimeter/Submillimeter Array (ALMA) and Atacama Pathfinder Experiment (APEX) in Chile, and the Institut de Radioastronomie Millimetrique 30m Telescope in Pico Veleta (PV).  Together, these present Earth-sized baselines to the primary EHT targets, Sgr A* and M87.

In April of 2017, the EHT successfully observed the black hole at the center of M87, and produced the first horizon-scale images of a black hole and its immediate surroundings.  With a mass and distance of $6.5\times10^9M_\odot$ and $16.8$~Mpc, the angular diameter of M87 is $42~\mu$as, and appears as an asymmetric ring with a deep central depression.  The flux asymmetry is suggestive of significant angular momentum, and the axis of angular momentum implied by the asymmetry is consistent with the jet axis at lower frequencies and larger scales \citep{EHT-4:19, EHT-5:19, EHT-6:19}. 

This jet is the most prominent feature of M87 beyond the horizon, extending to scales of 40~kpc \citep{Juno-Bire-Livi:99, Ly-Walk-Juno:07, Spar-Bire-Macc:96}.  The luminosity of M87's jet roughly $10^{44}~{\rm erg\,s^{-1}}$ as measured at a few different locations.  On kiloparsec scales observations of X-ray cavities inflated by the jet can be used to estimate jet power to about $10^{43}~{\rm erg\,s^{-1}}$ to $10^{44}~{\rm erg\,s^{-1}}$ with timescales of approximately 1 Myr \citep{Young-Wil-Mund:02}.  Below kiloparsec scales, superluminal optical features in the jet can be used to estimate the jet power, assuming they are the products of shocks.  The bright feature Knot A sits 0.9 kpc from the jet base and exhibits superluminal velocities up to 1.6c \citep{Meyer-Sparks-Biretta:13} and yields and estimated jet power of a few$\times10^{44}~{\rm erg\,s^{-1}}$ \citep{Bick-Begel:96} on a timescale of $10^{3}$ years.  The bright optical feature HST 1 sits 60 pc from the jet base, and contains superluminal features with velocities up to 6c \citep{Giro-Hada-etal:12}.  On timescales of 30 years, this feature yet again yields and estimated jet power of $10^{44}~{\rm erg\,s^{-1}}$ \citep{Sta-Aha-etal:06, Brom-Levin:09}.  

Prior VLBI observations have localized the jet to the near-horizon region \citep{Juno-Bire-Livi:99, Ly-Walk-Juno:07, Spar-Bire-Macc:96}.  Phased-reference observations, which permit VLBI-resolution astrometric observations, have verified that the relativistic jet photosphere converges on a single point at high frequencies \citep{Hada-Doi-etal:11}, with a jet width that decreases as a power-law with height, consistent with analytical expectations \citep{Blan-Koni:79}.  Previous EHT measurements of the size of the launching region have subsequently identified it unambiguously with the black hole \citep{Doel_etal:08, Doel_etal:12}.  

VLBI movies of M87 show clear evidence of time-dependent structure in the jet near the black hole \citep{Ly-Walk-Juno:07, Walk-Hardee-Davies:16, Hada-Kino-Doi:16}.  Multi-epoch imaging at 43~GHz over five years shows clear evidence of jet features with substantial proper motions.  A faint but visible counter-jet exists within $0.5~\mas$ of the jet core on the eastern side, with variable components moving away from the core at approximately 0.17c.  The western side of the jet is much brighter, with two clear limb-brightened arms on the north and south.  These arms contain smaller jet components with apparent velocities between 0.25c and 0.4c, but are most likely completely different components between observing epochs \citep{Ly-Walk-Juno:07}.  Similar observations  at 86~GHz also exhibit similar jet components, and with definite variability over 4 months of observations.  These higher frequency observations also see a weak, highly variable counter-jet within $0.25~\mas$ of the jet core, also with proper motions of approximately 0.17c.  A pair of limb-brightened arms extend westward of the jet core, with multiple jet components in both the north and south arm.  These components have apparent velocities between 0.1c and 0.48c, with proper motions approximately $1.0~\mas\,\yr^{-1}$ to $1.81~\mas\,\yr^{-1}$, with new components appearing on a timescale of a few months \citep{Hada-Kino-Doi:16}.  

The extreme luminosity of AGN jets are thought to originate from the conversion of gravitational potential energy to radiation from deep inside the the potential well of supermassive black holes.  The mechanism of this conversion is unconfirmed, though at present is believed to be facilitated by the extraction of rotational energy via magnetic torques, the chief distinguishing factor being the energy reservoir and topology of the magnetic fields.  These may be organized into two primary classes: jets driven by black holes \citep{BZ} and jets driven by accretion flows \citep{BP}.In the former, the jet is powered by the extraction of rotational energy via large-scale electromagnetic fields near the horizon of the supermassive black hole, which become highly collimated far from the black hole.
In the latter, the outflows are associated with a massive disk wind emanating from a hot accretion disk around the black hole, where the disk electromagnetic fields provide the motive force and the disk particles provides the luminous material.

In either case, the structure of the resulting jet is very similar.  The canonical jet model extracted from simulations features a force-free interior where the majority of the electromagnetic energy density is uncoupled from any particles in the jet \citep{McKi:06,Hawl-Krol:06,Tche-McKi-Nara:08}.  The exterior of the jet is composed of a magnetically dominated wind where the magnetic pressure is much larger than the gas pressure.  Both of these regions are, in turn, supported by a hot, thick accretion flow which provides the currents for the magnetic fields.

Simulated EHT images, which show sub-horizon scale emission structure, have been produced using a simplified, stationary, force-free model of the jet-interior \citep{Brod-Loeb:09}.  In this model, magnetic field lines that are anchored in a putative accretion flow, setting the boundary conditions for the force-free region.  Those field lines that anchor in the inner most stable circular orbit define a critical surface which exhibits the largest accelerations and highest Lorentz factors.  These field lines also serve as a convenient boundary of the jet and an important reference position when initializing our simulations, whose perpendicular distance from the jet axis we will call $\rho_{crit}$.  

Simulated images from semi-analytic models differ from images generated from general relativistic magnetohydrodynamic (GRMHD) simulations primarily through the distributions of the synchrotron emitting lepton populations.  While the magnetic field and energy density are well described in GRMHD simulations, the particle content in regions of high magnetic density is not well recovered \citep{DMA11}.  Non-thermal particle emission in the highly magnetized regions could arise from turbulence or magnetic reconnection events well below the simulation resolution, and efforts to produce a self-consistent model for the evolution of the non-thermal particle distribution are still underway \citep{MosFalShio:16}.  Because of this, jet variability in these GRMHD simulations are strongly dependent on numerical noise in the jet-launching region and may not adequately model observed variability in M87.  Indeed, when comparing GRMHD models to the observations of M87, the emissivity was set to zero in regions where the magnetization was order unity due to the unreliable nature of the emission \citep{EHT-5:19}.  

For simplicity, the force-free jet model used in this work is time-stationary, and we add variability by introducing compact emission regions within the jet-launching zone, but do not allow these compact regions to affect the field structure of the jet.  As a result some intrinsic variability that is expected from the turbulent accretion flow and jet is not captured by the semi-analytic model, but we also do not impose any restrictions on the particle emissivities in regions of high magnetization.  Force-free jets are not generally time stationary, but this work serves as a pilot study to explore the effects of hot-spot variability in the jet launching region as a way to distinguish black hole driven and wind-driven jets.  While outside the scope of this work, one could physically motivate these spots as reconnection products at the jet boundary, particle creation events inside the jet, or as wind material blown up from the accretion disk.  These ``spots'' are similar to those employed within the accretion flow environments in \citet{Brod-Loeb:06b}, with two key exceptions: 1.~following injection they subsequently outflow along the highly-stratified velocity field of the stationary force-free jet model, and 2.~they necessarily shear.  We find that both the small- and large-scale features of the 1.3~mm images are strongly dependent on the initial spot injection location.  Thus, the nature and properties of the variability in EHT observations of M87 is diagnostic of the location of the jet launching region and the origin of the emitting jet leptons.

In Section 2 we describe how spots are modeled in detail.  In Section 3 we explore a variety of different injection sites and spot parameters, identifying qualitative trends and quantitative signatures of black hole-driven and disk-driven jets.  Conclusions are collected in Section 4.

Throughout the paper we specify distances in $M\equiv GM/c^2 = 1.0\times10^{15}~\cm$ and time in $t_g\equiv GM/c^3 \approx 9~\hr$.  Where appropriate, values in other units will be provided for convenience.  In all models, the black hole has a dimensionless spin of $a\equiv J/M=0.993$, where $J$ is the spacetime angular momentum.  While the M87 images and model comparisons produced by the EHT were not able to constrain the spin of M87, the jet power at larger scales and other frequencies are difficult to reproduce in GRMHD simulations with low spin \citep{EHT-5:19}.  All images are simulated observations at 1.3~mm, and show a logarithmic intensity scale over four orders of magnitude, normalized so that the total image intensity of just the quiescent jet without a spot is about $0.7~\Jy$.

\section{Modeling the Variability}

\subsection{Non-shearing Spot}

The radio wavelength images of M87 made by \citet{Ly-Walk-Juno:07, MLWH-M87:16, Hada-Kino-Doi:16}, among others, show clear evidence of structure in the jet at lower frequencies.  While this work does not seek to explain the physical mechanisms of generating these structures, we do want to explore how the EHT might see jet structure near the black hole.  To this end, we model these structures, or spots, as an over-density of non-thermal electrons which we construct to initially have a spherical Gaussian distribution in the reference frame of the spot center.  The physical motivation for these spots is that a high energy event occurs near the base of the jet near the black hole, generating a large population of non-thermal particles.  The jet is overwhelmingly dominated by Poynting flux this close to the black hole \citep{Tche-McKi-Nara:08,MTB12}, so large fluctuations in the emissivity should not significantly alter the larger jet structure.  The motion of the spot is determined solely by the location of the spot center, which follows a single stream line up the jet, and the density profile of the spot stays fixed to the spot center.

\begin{figure}
 \includegraphics[width=\columnwidth]{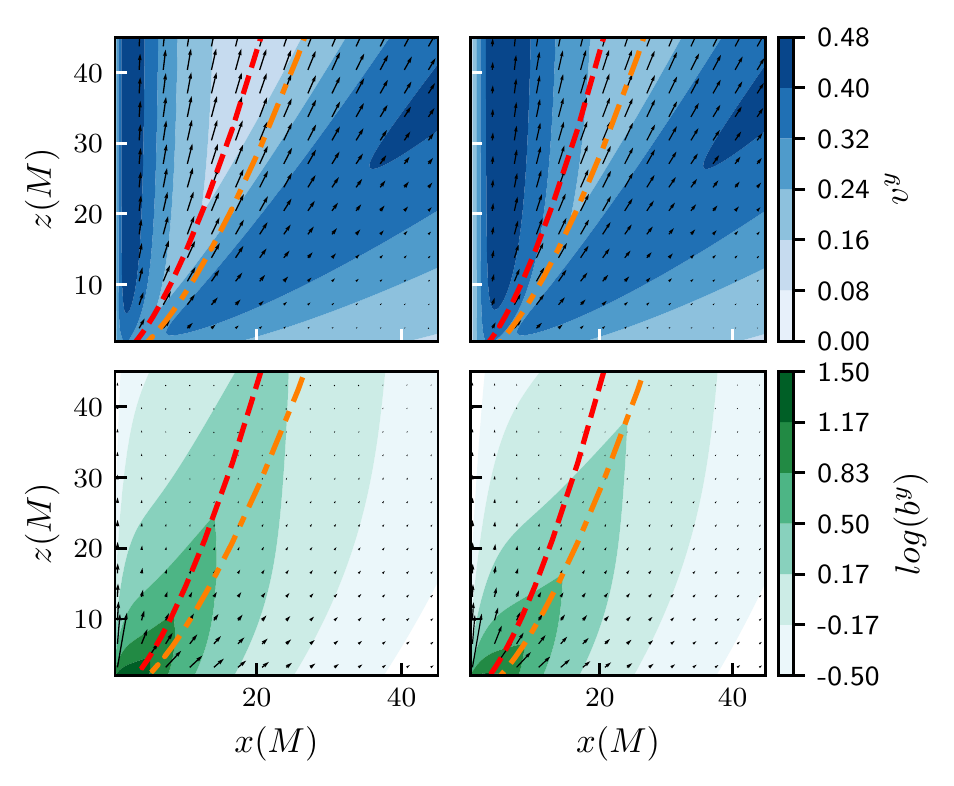}
 \caption{Velocity (blue) and Magnetic (green) fields of the jet in the x-z plane.  The vector field shows the magnitude and direction of the $x$ and $z$ (poloidal) components, and the contours show the magnitude of the $y$ (toroidal) component, where darker contours correspond to higher values.  As we travel up the jet, the magnitude of the toroidal velocity component ($v_{y}$ in this slice) drops, while the poloidal velocity component ($v_{x}$ and $v_{z}$) increases.  The dashed curves represent the streamline connected to the inner jet edge.  The red dashed curve is the streamline for an inner jet edge at the ISCO, and the orange dashed curve is the streamline for an inner jet edge at $2 ~r_{ISCO}$.}
 \label{fig:VelField}
\end{figure}

Our jet model produces a velocity field that accelerates particles up the jet, but also includes significant shearing in the azimuthal plane.  In Figure \ref{fig:VelField}, we show the velocity (blue) and magnetic (green) field structures.  In our jet model, we fix the angular velocity $\Omega$ inside the critical surface (red dashed line) to the value at the critical surface, as described in \citet{Brod-Loeb:09}.  In many simulations, $\Omega$ is generally smaller than the angular velocity at the ISCO, as in \citet{MTB12}.  To investigate the effects of reduced $\Omega$ on the velocity field, we artificially extended the jet inner radius away from the ISCO (orange dashed line).  This serves to decrease $\Omega$ inside the critical surface.   As shown in Figure \ref{fig:VelField}, increasing the jet inner edge reduces the angular velocity (blue contours) inside the critical surface, and produces a shallower $\Omega$ gradient across the cylindrical radius.  Outside the critical surface, $\Omega$ is unchanged.  Thus, reducing $\Omega$ inside the critical surface should have no effect on hotspots launched outside the critical surface, and will only affect the evolution of spots launched inside the critical surface.  Regardless, extended emission regions still become distorted and sheared out, changing the emission profile dramatically depending on where in the velocity field the spot is launched.  A spot with a fixed density profile will not be able to adequately simulate expected variability in our jet velocity field, so we must develop a method that allows the spot to shear out while maintaining a good approximation of the initial density profile of our non-shearing spot.  

\subsection{Shearing Spot}

While our Gaussian spots should adequately model the general motion of over-densities in the jet, the shearing forces near the black hole non-trivially affect the structure of these spots as they evolve.  In order to confidently explore the impact of this type of jet variability on EHT images, we need to think carefully about how to initially construct and keep track of the changing spot.  We construct a shearing spot as an assembly of smaller non-shearing spots arranged in equal mass shells that approximate the density structure of the non-shearing spot.  These mini-spots serve as Lagrangian control points for calculating the local density in a way similar to smoothed particle hydrodynamics methods.  

To be able to propagate the mini-spots through the jet velocity field, we need a set of coordinates $\xi$ that co-move with each mini-spot.  
\begin{eqnarray}
\xi_1^\mu&=&\epsilon^{\mu\alpha\beta\gamma}t_{\alpha} z_{\beta} u_{\gamma} \\
\xi_2^\mu&=&\epsilon^{\mu\alpha\beta\gamma}t_{\alpha} \xi_{1\beta} u_{\gamma}  \\
\xi_3^\mu&=&\epsilon^{\mu\alpha\beta\gamma}\xi_{1\alpha} \xi_{2\beta} u_{\gamma},
\end{eqnarray} 
where $\epsilon^{\mu\alpha\beta\gamma}$ is the 4th-dimensional Levi-Civita symbol, $t_{\alpha}$ is the lab frame time-like Killing vector, $z_{\beta}$ is the lab frame vertical spatial unit vector, and $u_{\gamma}$ is the lab frame velocity.  Once these $\xi$ are normalized, they serve as a space-like orthonormal triad that is also orthogonal to the lab frame velocity; together they form a complete spacetime basis.

\begin{figure}
  \begin{center}
    \includegraphics[width=\columnwidth]{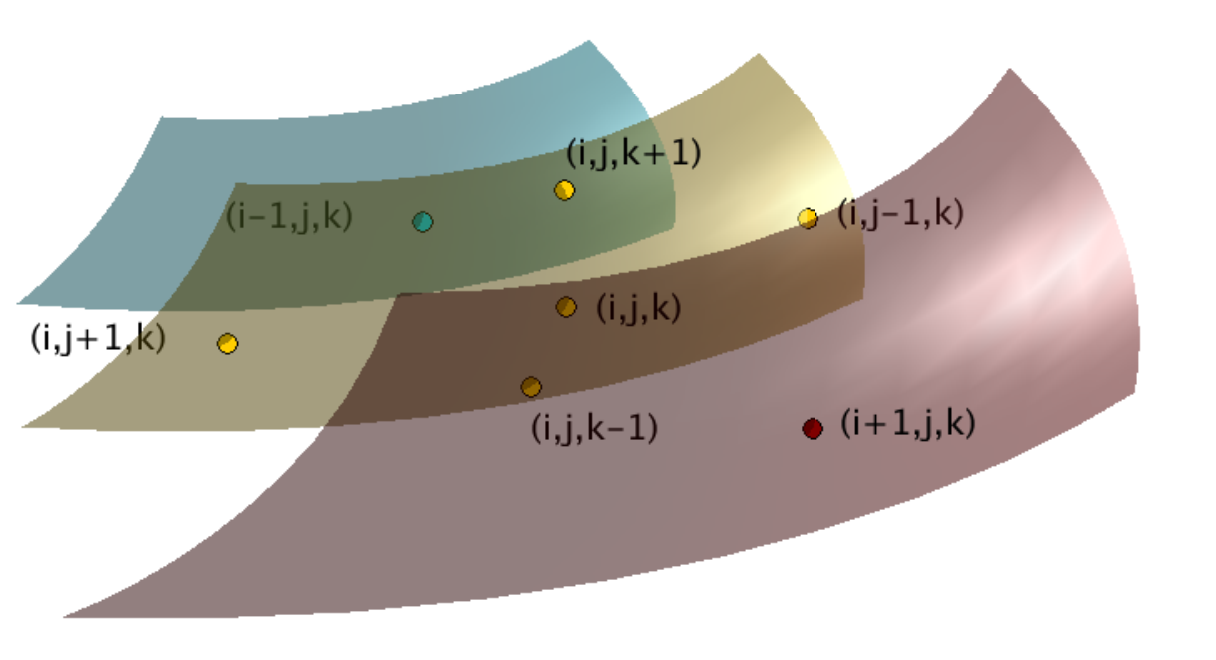}
  \end{center}
  \caption{Sections from the outermost radial mini-spot shells showing central mini-spot $(i,j,k)$ and its nearest neighbors in the cardinal shell directions.  The mini-spots have been interpolated $5t_g$ from their initial launch time, and significant shearing from square in the $\theta$ and $\phi$ directions are apparent.}
\label{fig:Shells}
\end{figure}

We divide our non-shearing density profile into radial, azimuthal, and polar shells evenly distributed by mass.  From our co-moving cartesian basis we then construct the initial mini-spot distribution in Boyer-Lindquist coordinates $x_{ijk}^\mu$ where each $i$, $j$, and $k$ index corresponds to a single mini-spot shell.
\begin{equation}
\begin{aligned}
x_{ijk}^\mu &= \rho_i \sin(\vartheta_j) \cos(\varphi_k) \xi_1^\mu+\rho_i\sin(\vartheta_j)\sin(\varphi_k)\xi_2^\mu \\
&\qquad+ \rho_i\cos(\vartheta_j)\xi_3^\mu + r^\mu,
\end{aligned}
\end{equation}
where $\rho$, $\vartheta$ and $\varphi$ are the shell indices for each mini-spot, and $r^\mu$ is the initial spot center position in the lab frame.  Every mini-spot appears simultaneously in the frame co-moving with the spot center, but will appear at different Boyer-Lindquist times in the lab frame.

To find the mini-spot positions at later times, we calculate each mini-spot trajectory $dx_{ijk}^{\mu}/dt$ using the value of the velocity field $u_{ijk}^{\mu}$ at the mini-spot,
\begin{equation}
\frac{dx_{ijk}^{\mu}}{dt} = \frac{u_{ijk}^{\mu}}{u_{ijk}^{0}}.
\end{equation}
We use a 4th/5th order Cash-Karp Runge-Kutta method to integrate along these trajectories, with adaptive step sizes determined by the error of the 4th order calculation \citep{NumRecipes3:07}. We first integrate backwards from our initial Boyer-Lindquist time to make sure we have mini-spot trajectories for every mini-spot in the lab frame, then integrate forward along the trajectories from the initial time to far enough in the future to travel significantly up the jet.  We smoothly propagate each control point along their own streamline in the jet velocity field to create a spot path, which we tabulate at each integration step.

\begin{figure}
  \includegraphics[width=0.49\columnwidth]{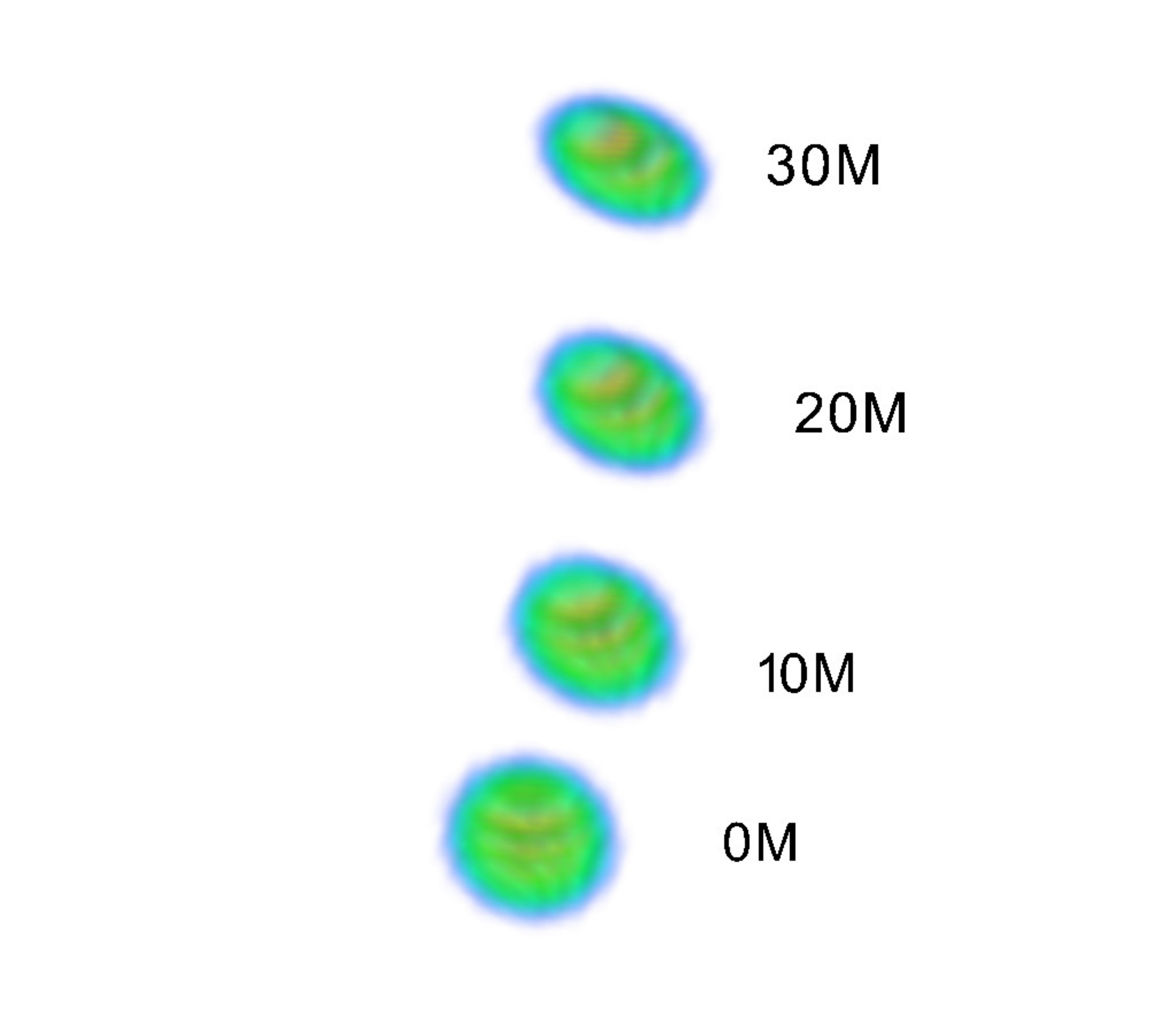} 
  \includegraphics[width=0.49\columnwidth]{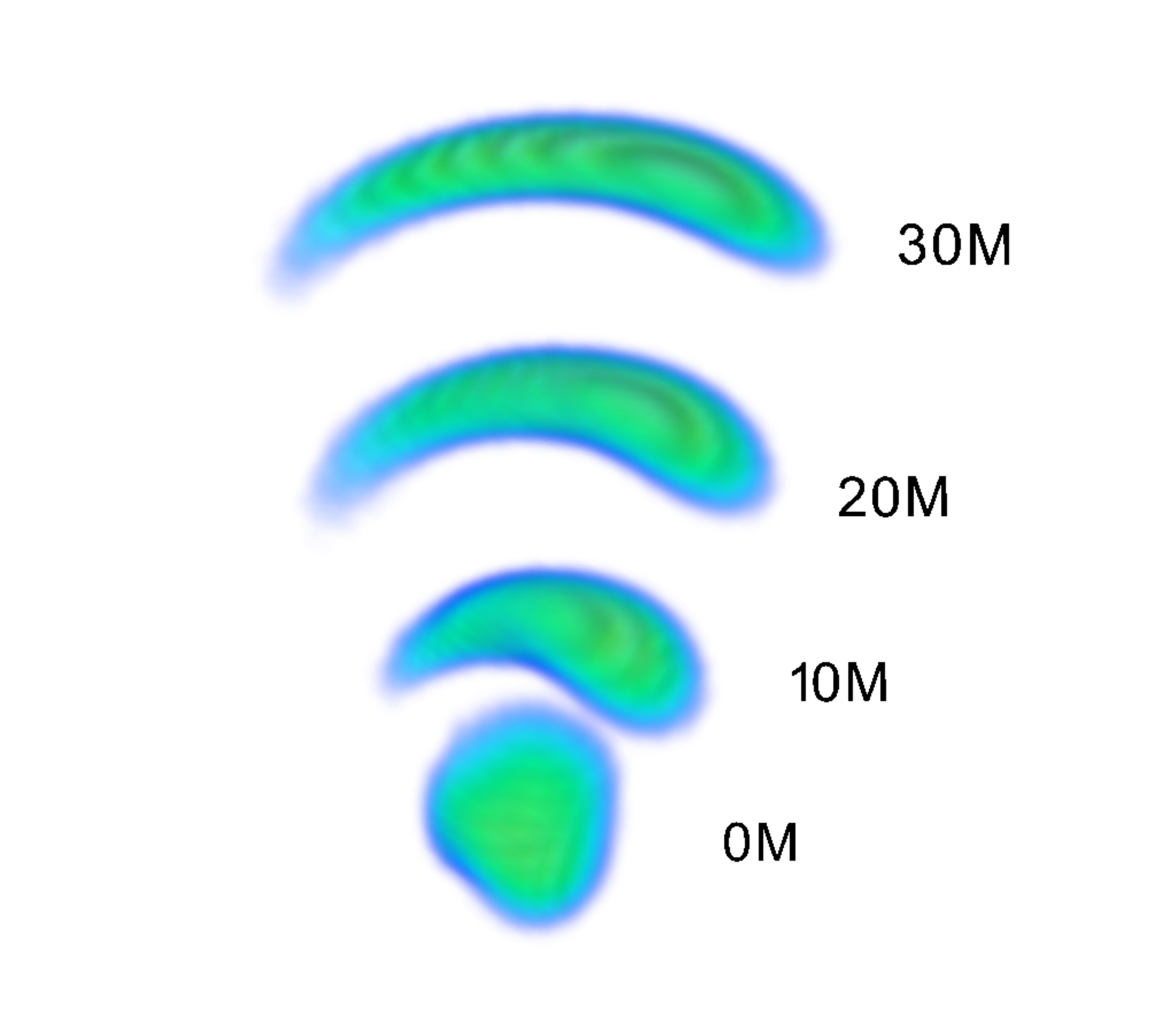} 
  \caption{Density evolution of a non-shearing and shearing spot launched at $\rho=0.5 \rho_{\rm crit}$ and $\phi=270^{\circ}$ for $30t_g$ in gravitational time.  The top set of spots are a non-shearing spot with a Gaussian density profile in the spot-center reference frame, and the bottom set of spots are a shearing spot with the same initial launch position and density profile.  The shearing spot develops an extended tail within a few $t_g$ of launch, producing a very different density profile than the non-shearing spot.}
  \label{fig:Devol}
\end{figure}

We can calculate the density at any point by adding up the density contribution from all the mini-spots nearby:
\begin{equation}
n_s = \frac{\rho_0 r^{3}_{s}}{N_s}\sum_{j}\frac{\exp{\left(-\frac{1}{2}\Delta r_j I_{j}^{-1}\Delta r_j\right)}}{\sqrt{| I_{j}^{-1} |}}.
\end{equation}
Our total density $n_s$ is made up of $j$ mini-spots with individual number densities normalized by the correlated distance between one mini-spot and its nearest neighbors.   Here, $\rho_0$ is the number density for an equivalently sized non-shearing spot, $r_s$ and $N_s$ are the spot radius and total number of mini-spots respectively, and $\Delta r_j$ is the distance from where we want to calculate the density, at $(r, \theta, \phi)$, to the mini-spot position $x_{ijk}^\mu$.  We add up the density contributions from each mini-spot by using an inverse covariance matrix $I$ of mini-spot positions, which lets us calculate the local density in the mini-spot coordinate frame:
\begin{eqnarray}
 \label{eq:i1}
 I_{rr} &= \sum_{[ i' j' k']} \left(r_{i'j'k'}-r_{ijk} \right)^2  \\
 \label{eq:i2}
 I_{\theta\theta} &= \sum_{ [i' j' k']} \left(\theta_{i'j'k'}-\theta_{ijk} \right)^2  \\
 \label{eq:i3}
 I_{\phi\phi} &= \sum_{[i' j' k']} \left( BC\left( \phi_{i'j'k'}-\phi_{ijk} \right) \right)^2   \\
 \label{eq:i4}
 I_{r\phi} &= \sum_{[i' j' k']} \left(r_{i'j'k'}-r_{ijk} \right) \left( BC\left(\phi_{i'j'k'}-\phi_{ijk} \right) \right).
\end{eqnarray}
The sums in Equations (\ref{eq:i1} -- \ref{eq:i4}) are over the mini-spot nearest neighbors to the central $(i,j,k)$ location (in spot coordinates) where we want to calculate the density.  The nearest neighbor spots are relative to their shell coordinates; the nearest $r$ neighbors live in the $i+1$ and $i-1$ shells, the nearest $\phi$ neighbors live in the $j+1$ and $j-1$ shells, and the nearest $\theta$ neighbors live in the $k+1$ and $k-1$ shells.  The $BC$ function is a branch cut function that makes sure we calculate the shortest distance between two $\phi$ shells, to make sure that mini-spots at $\phi=355^{\circ}$ and $\phi=5^{\circ}$ are properly separated by .
We only consider the covariance of nearest neighbors in the $r-r$, $\phi-\phi$, $\theta-\theta$, and $r-\theta$ directions because we do not expect much shearing in the $\theta-\phi$ or $r-\phi$ directions.  Indeed, it is apparent in Figure \ref{fig:Shells} that this is the case, with the most significant shearing occurs in the radial and poloidal directions.  Finally, the proper density is given by $n_s/u^t$.

\begin{figure}
  \includegraphics[width=0.49\columnwidth]{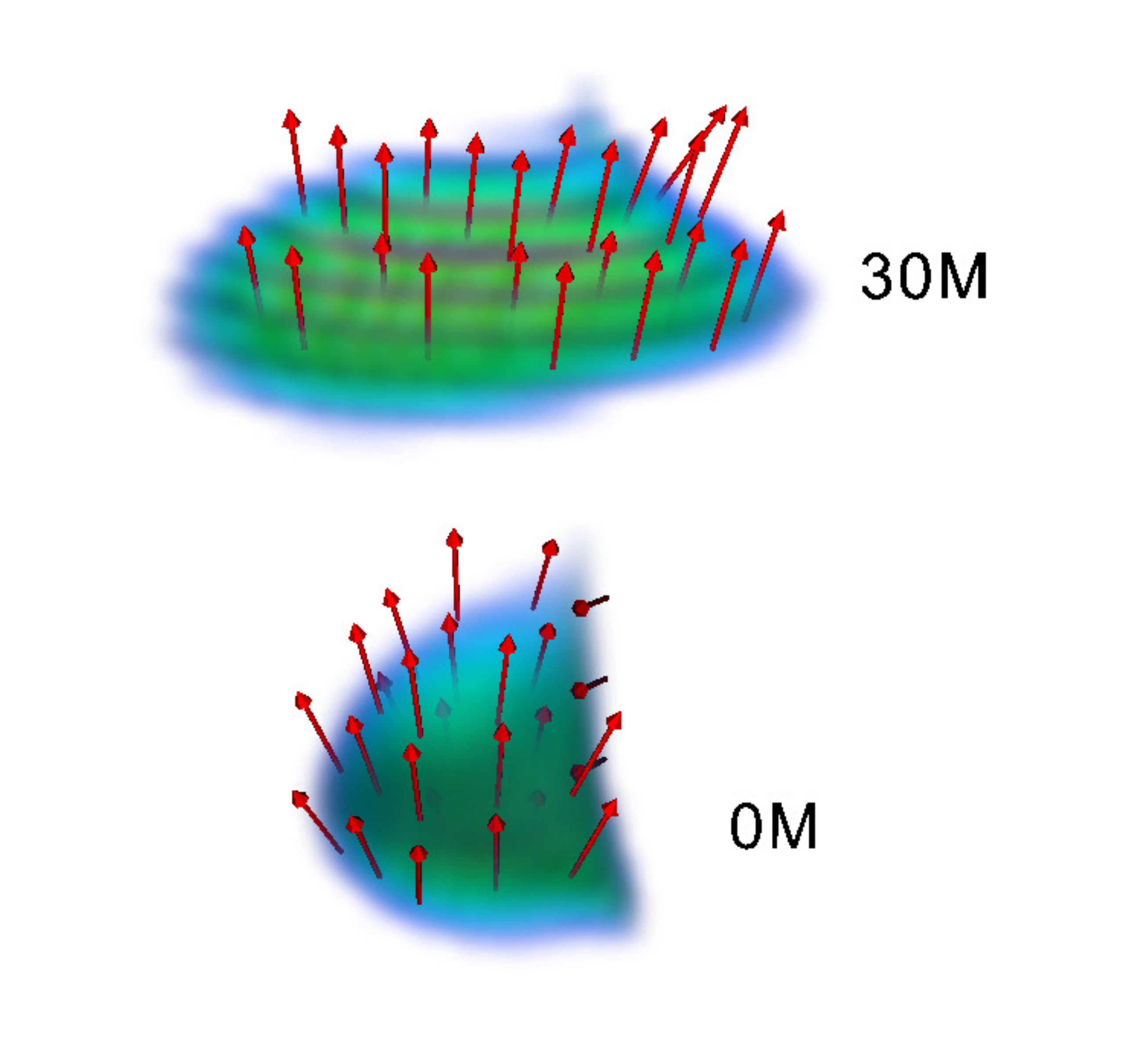}    
  \includegraphics[width=0.49\columnwidth]{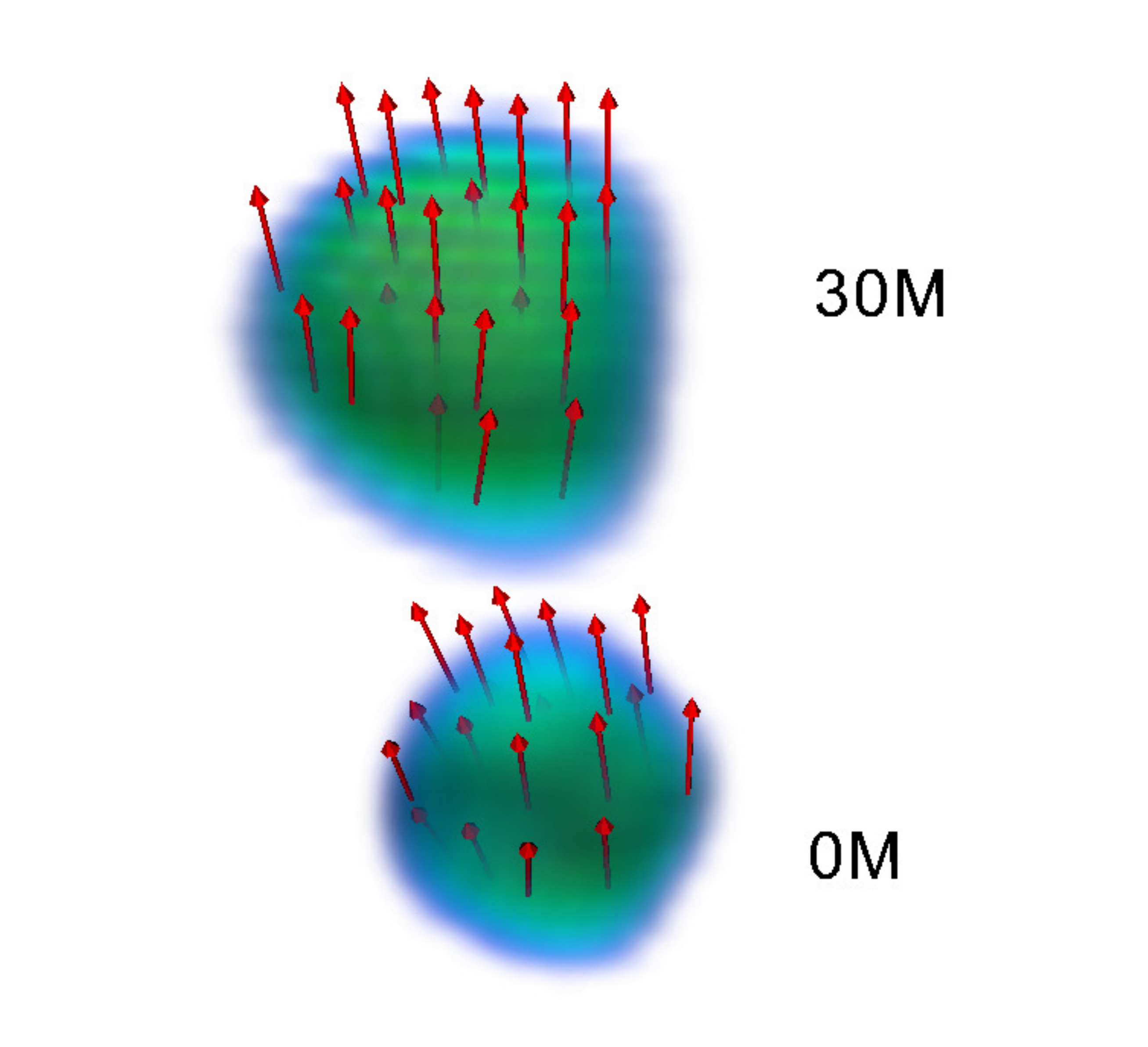} \\
  \caption{Density and velocity profiles for the spot launched at $\rho=0.75 \rho_{\rm crit}$ (left) and $1.25 \rho_{\rm crit}$ (right) at $t=0t_g$ (bottom) and $30t_g$ (top).  The velocity profile for the outer spot (bottom two blobs) exhibits much less shearing, resulting in a more spherically symmetric spot as it evolves than the interior spot.}
  \label{fig:DexEvol}
\end{figure}

Black hole driven spots should have a much more extended density profile than a non-shearing Gaussian spot launched with the same initial conditions.  This is apparent in Figure \ref{fig:Devol}, which shows the shearing spot with an extended tail feature forward of the spot central density, and a much more extended density region in general.  Shearing spots launched at different launch radii should also develop different density profiles, demonstrated in Figure \ref{fig:DexEvol}, since the velocity field structure changes significantly outside the critical field surface.  

Even though spots might only differ in initial radial launch distance by less than $5 r_g$ they exhibit substantially different density profiles after even a few gravitational time steps.  A black hole driven spot will experience a much stronger velocity gradient than a wind-driven spot, and should be distinguishable in much the same way a non-shearing Gaussian spot is distinguishable from a shearing spot.  Even though our jet model is relatively simple compared to GRMHD simulations, the techniques used to generate and track hotspots spots can be used in other jet models with more complicated velocity fields.  

In practical terms, the shearing spot is initialized in terms of cylindrical launch coordinates, centered on the black hole.  Throughout the rest of this paper, we will reference the radial cylindrical coordinate $\rho$, not to be confused with the initial non-shearing spot density profile.  This radial launch coordinate is parametrized in units relative to the cylindrical radius of the critical field line or surface, which we denote as $\rho_{crit}$.  Additionally, we can initialize the spot at different azimuthal positions around the black hole, which we denote with $\phi$.  With respect for the line of sight, a spot launched in front of the jet axis is initialized with $\phi=0^{\circ}$, and a spot launched behind the jet axis is initialized with $\phi=180^{\circ}$.  This should also not be confused with the inclination of the jet, which is fixed in our simulations to be $i=17^{\circ}$, consistent with the results from \citet{MLWH-M87:16}.  The vertical position can also be specified, but is set to be $z=3M$ for most of the simulations described in this paper.  We discuss modifications to the launch height in Section 3.5.

\section{Exploration of Imaging}

\subsection{Radiative Transfer}

The primary emission mechanism near the black hole is synchrotron radiation from thermal and non-thermal electrons.  The thermal component is modeled using the emissivity described in \citet{yuan03}, and corrected to be fully covariant following the description of covariant radiative transfer by \citet{Brod-Blan:04}.  We assume the distribution of thermal electrons is isotropic, and use a thermal synchrotron polarization fraction derived in \citet{petrosian83}.  

\begin{figure}
  \includegraphics[width=\columnwidth]{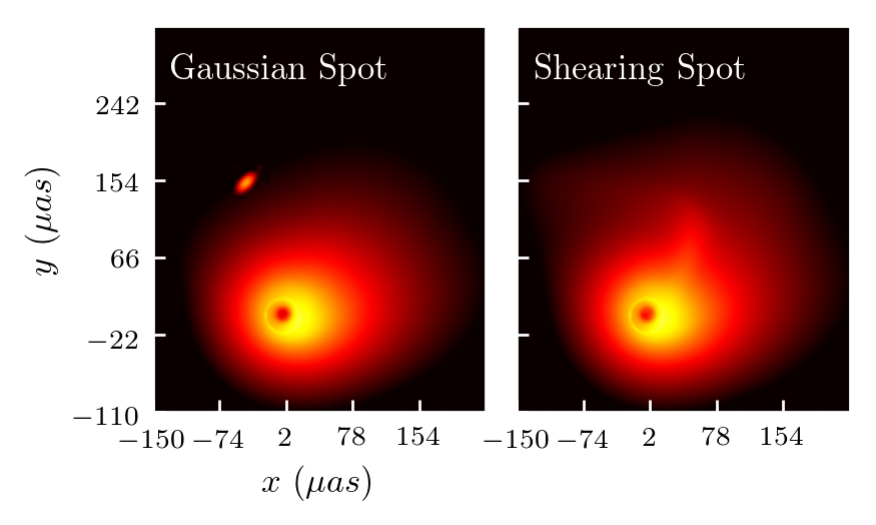}   
  \caption{Snapshots at late times ($35t_g$) of a Gaussian and Shearing spot launched at $\rho=0.5 \rho_{\rm crit}$ and $\phi=270^{\circ}$.  The non-shearing spot stays compact and bright, but the shearing spot develops complicated emission structures in the image plane, due to a combination of light delays and more spatially extended emission.}
  \label{fig:gexcomp}
\end{figure}

\begin{figure*}
  \begin{center}
   \includegraphics[width=\textwidth]{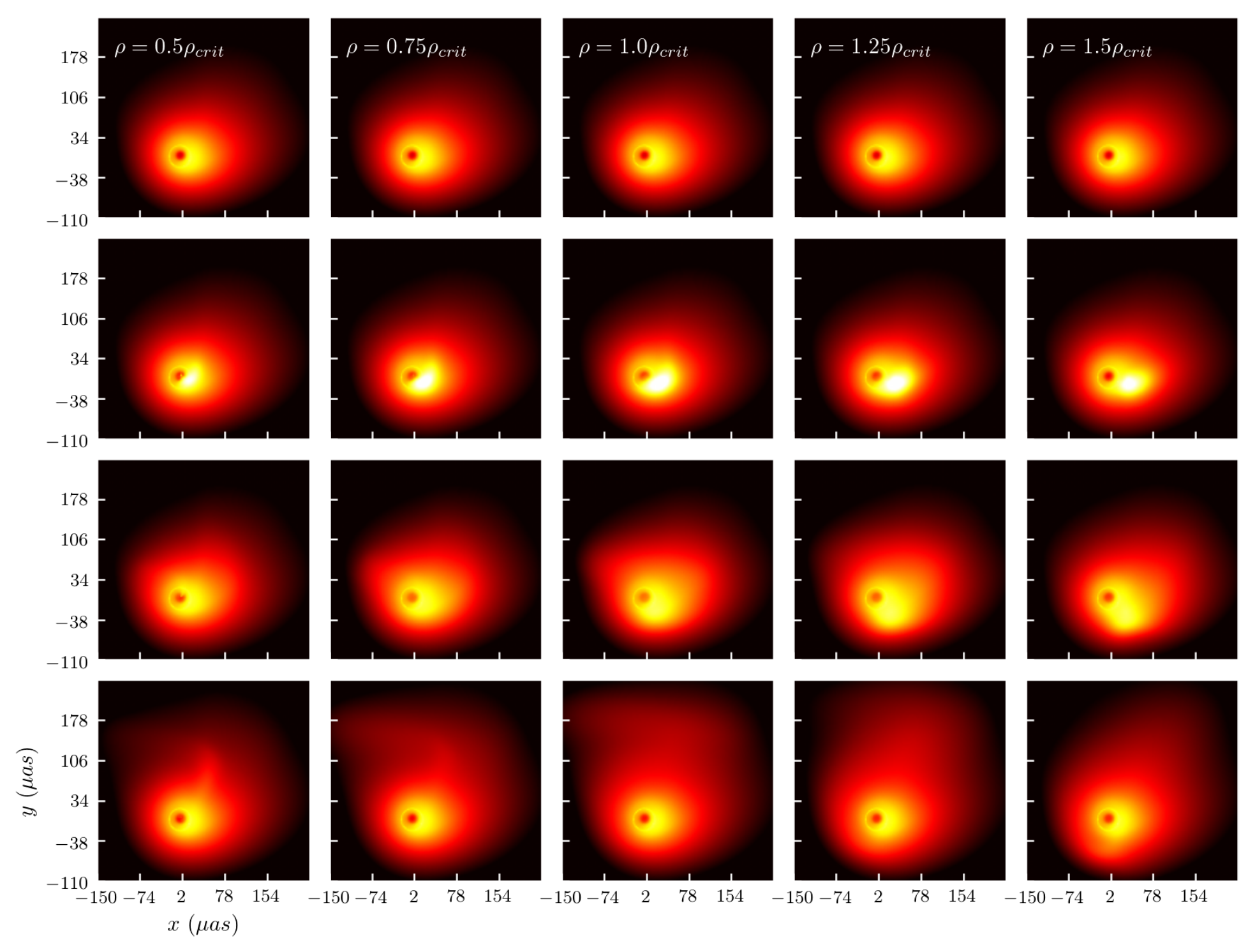} 
  \end{center}
  \caption{
    Snapshots of a shearing spot launched at different cylindrical positions, both inside and outside of the jet, showing evolution over time.  From left to right, the columns correspond to spots with initial cylindrical radii at $0.5\rho_{\rm crit}$, $0.75\rho_{\rm crit}$, $\rho_{\rm crit}$, $1.25\rho_{\rm crit}$, and $1.5 \rho_{\rm crit}$.  In all cases the azimuthal launch location was $\phi=270^{\circ}$.  Time flows from the top row to the bottom row, where between each row the observer time advances by $10t_g$ (90~hr).  Spots launched inside the critical surface exhibit more complex emission structures when compared to spots launched outside the critical surface.}
  \label{fig:exradzoom}
\end{figure*}

The non-thermal emission is assumed to follow a power-law distribution which cuts off below some critical Lorentz factor, as described by \citet{jones77}.  The cut-off is fit to match M87's observed milimeter spectrum \citep{Brod-Loeb:09}, and the absorption coefficients are determined directly from Kirchoff's law \citep{Brod-Blan:04}.

\subsection{Shearing v. Non-Shearing Spots}

The distributed density profile of the shearing spots is easily distinguishable from the Gaussian spot in the imaging space as well.  
It is readily apparent that the shearing spot has a much more extended intensity profile.  This Figure \ref{fig:gexcomp} shows a snapshot at $35 t_g$ (about 315 hours or 13 days) after the initial launch, and the shearing spot exhibits a complex structure.  The spot has sheared out enough to wrap up on itself at least once, creating a distinct arc over the bright jet emission near the black hole.  

\subsection{Shearing Spots, Changing Radius}

We have demonstrated that the shearing spot is substantially different from the Gaussian spot in both its density profile and in the image domain.  Our next task is to compare shearing spots at different launch positions.  The M87 model used in this paper has a large spin, $a=0.993$, resulting in a strong velocity field very close to the black hole. This means photons from our spots can be strongly beamed towards or away from us when the spot is near the black hole, in addition to beaming associated with the acceleration from the jet itself.  The geometry of the velocity field near the jet will also lead to different shearing morphologies, depending on whether the spot was launched inside or outside the jet critical surface.  

\begin{figure*}
  \begin{center}
   \includegraphics[width=\textwidth]{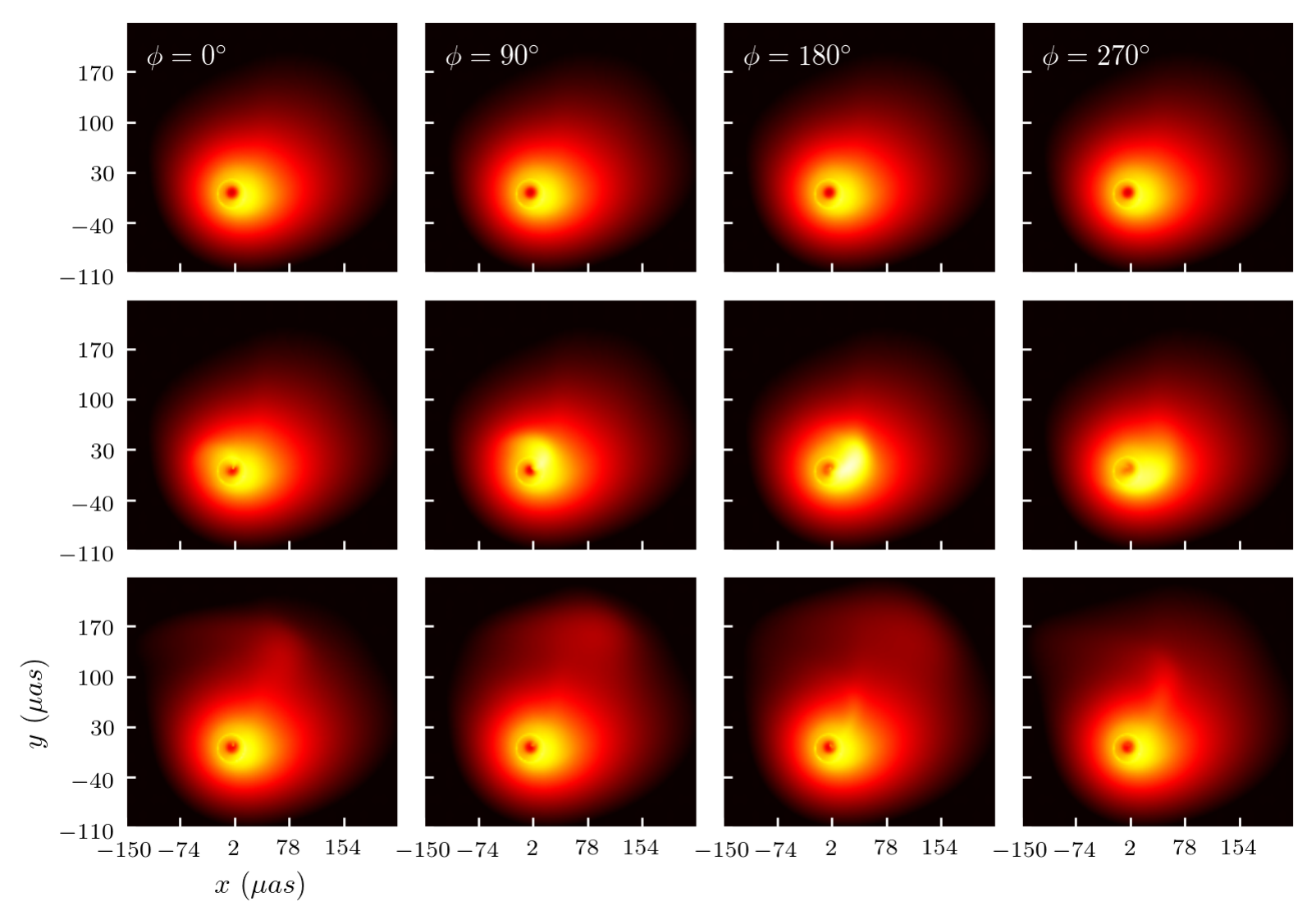} 
  \end{center}
   \caption{Images of a shearing spot at $\rho=0.5\rho_{\rm crit}$ at cylindrical angles $\phi$ of $0^{\circ}$, $90^{\circ}$, $180^{\circ}$ and $270^{\circ}$ showing early evolution.  Time flows from the bottom row to the top in steps of $15 t_g$ (5 days), starting $5 t_g$ after the spot is launched.  Changing the launch azimuthal position dramatically changes the spot intensity and location of extended structure in the image.}
  \label{fig:ex50AzZoom}
\end{figure*}

Zooming into the immediate region around the black hole at early times, we can still see differences in image structure when we change the spot launch radius.  Even though the spot is very bright at early times, we can see in the top two rows of Figure \ref{fig:exradzoom} that black hole-driven spots shear away faster than wind-driven spots.  We can also see how spots starting at the critical radius initially look very much like black hole-driven spots, but shear into structures similar to wind-driven spots at later times.  

Spot densities for wind-driven spots remain relatively compact, as seen above in Figure \ref{fig:DexEvol}.  Even after $30 t_g$ (11 days), wind-driven spots are only slightly more extended compared to their launch size.  Nevertheless, propagation time delays smear the spots on the image plane.  While this happens for black hole-driven spots as well, this phenomenon is easiest to see in wind-driven spots, where the spot both shears and travels much more slowly.  

If we zoom out and look at the spot for longer, as we do in the bottom two rows of Figure \ref{fig:exradzoom}, the same conclusions still apply: azimuthal shearing is most significant inside the critical surface (about $4 r_g$ for a launch height of $3 r_g$), which is responsible for low surface brightness arcs.  Outside the critical surface, azimuthal shearing is significantly suppressed, and the spot slowly expands and flows outward, generating an extended arm in the last column.

Figure \ref{fig:exradzoom} demonstrates that spots launched inside the jet critical surface exhibit sheared structures that distinguish them from spots launched outside the critical surface.  The spot launched at $\rho=0.5\rho_{\rm crit}$ (first column) looks much different than the spot launched at $\rho=1.5\rho_{\rm crit}$ (last column).  The black hole-driven spot creates a thin arc in the last row, $35 t_g$ (13~days) after the spot was launched.  The spot arc is still present for the spot launched on the critical surface, whereas the exterior spots have a more extended emission region coming straight out of the bright jet region.  Spots launched outside the critical surface exhibit much dimmer arcs, and perpendicular structures disappear entirely for spots launched only $2 r_g$ ($\rho=1.5\rho_{\rm crit}$) away from the critical surface. 

\begin{figure*}
  \begin{center}
   \includegraphics[width=\textwidth]{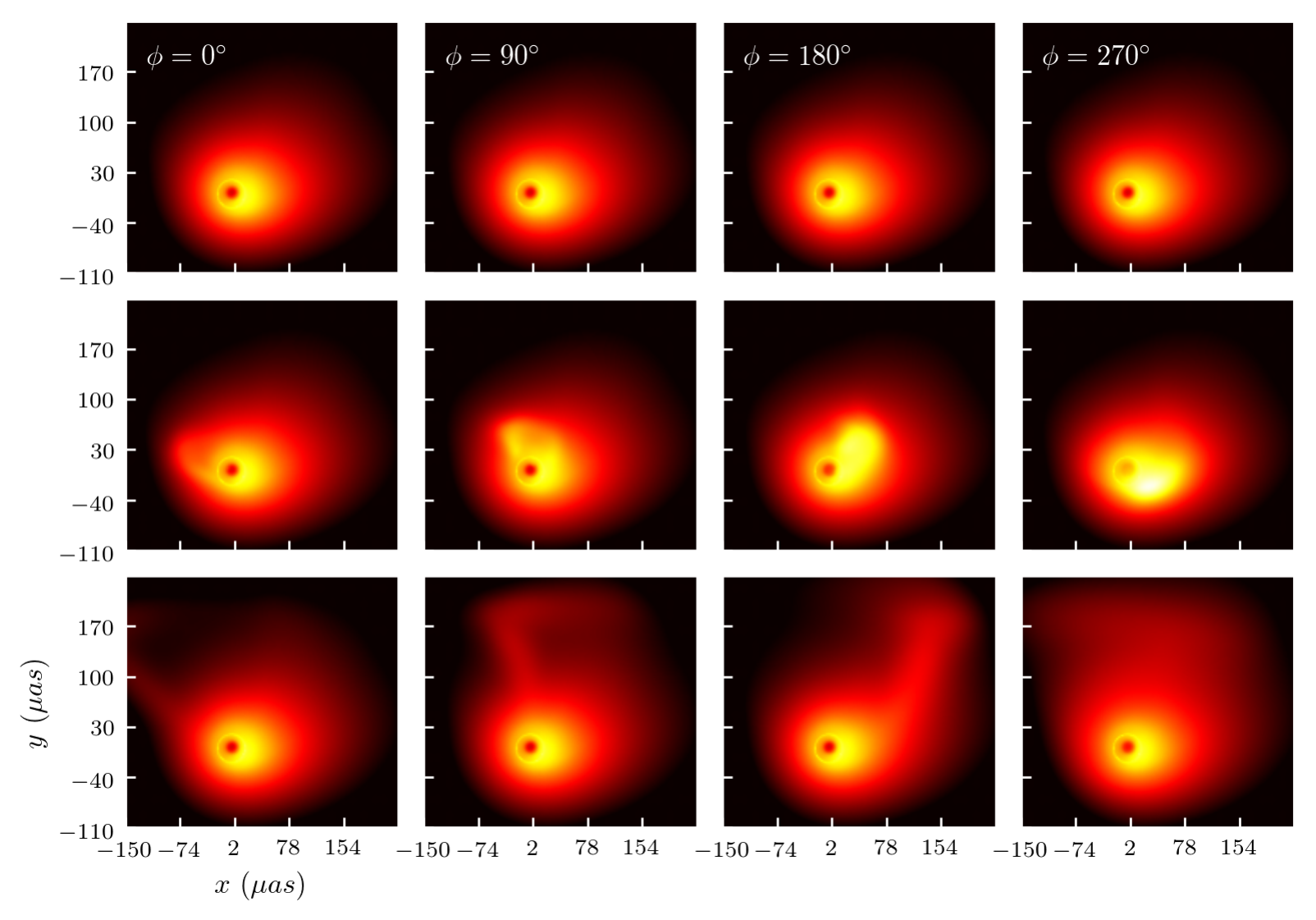} 
  \end{center}
  \caption{Images of a spot launched at $\rho=1.0\rho_{\rm crit}$ with $\phi$ angles of $0^{\circ}$, $90^{\circ}$, $180^{\circ}$, and  $270^{\circ}$ showing early evolution.  The top row is $5 t_g$ (45 hours) after the spot launch, the second row is $20 t_g$ (7.5 days) after the spot launch, and the bottom row is $35 t_g$ (13 days) after the spot launch.  Changing the azimuthal launch parameter produces significant differences in image intensity, with spots launched at low angles contributing to a much lower image intensity than spots launched at large angles.  Complicated arc structures away from the black hole are evident at late times. }
  \label{fig:ex100AzZoom}
\end{figure*}

Without a spot, the jet electron density is set such that the total intensity for a quiescent jet is approximately 0.7~Jy, which is consistent with the measurements of compact flux from geometric model fits in \citet{EHT-6:19}.  For a fixed spot electron density, the maximum intensity in the image depends strongly on the initial radial launch position.  The spot appears brightest when launched on the critical surface, and falls off faster when launched inside the jet.  The shorter spot half-life for black hole driven spots can be attributed to stronger shearing compared to wind driven spots, which serves to disrupt and dilute the spot intensity inside the jet.  Spots outside the jet experience less shear, and remain relatively more compact for longer.  A more in depth discussion about spot light curves is reserved for Section 3.6

\subsection{Changing Launch Azimuth}

\begin{figure*}
  \begin{center}
   \includegraphics[width=2.1\columnwidth]{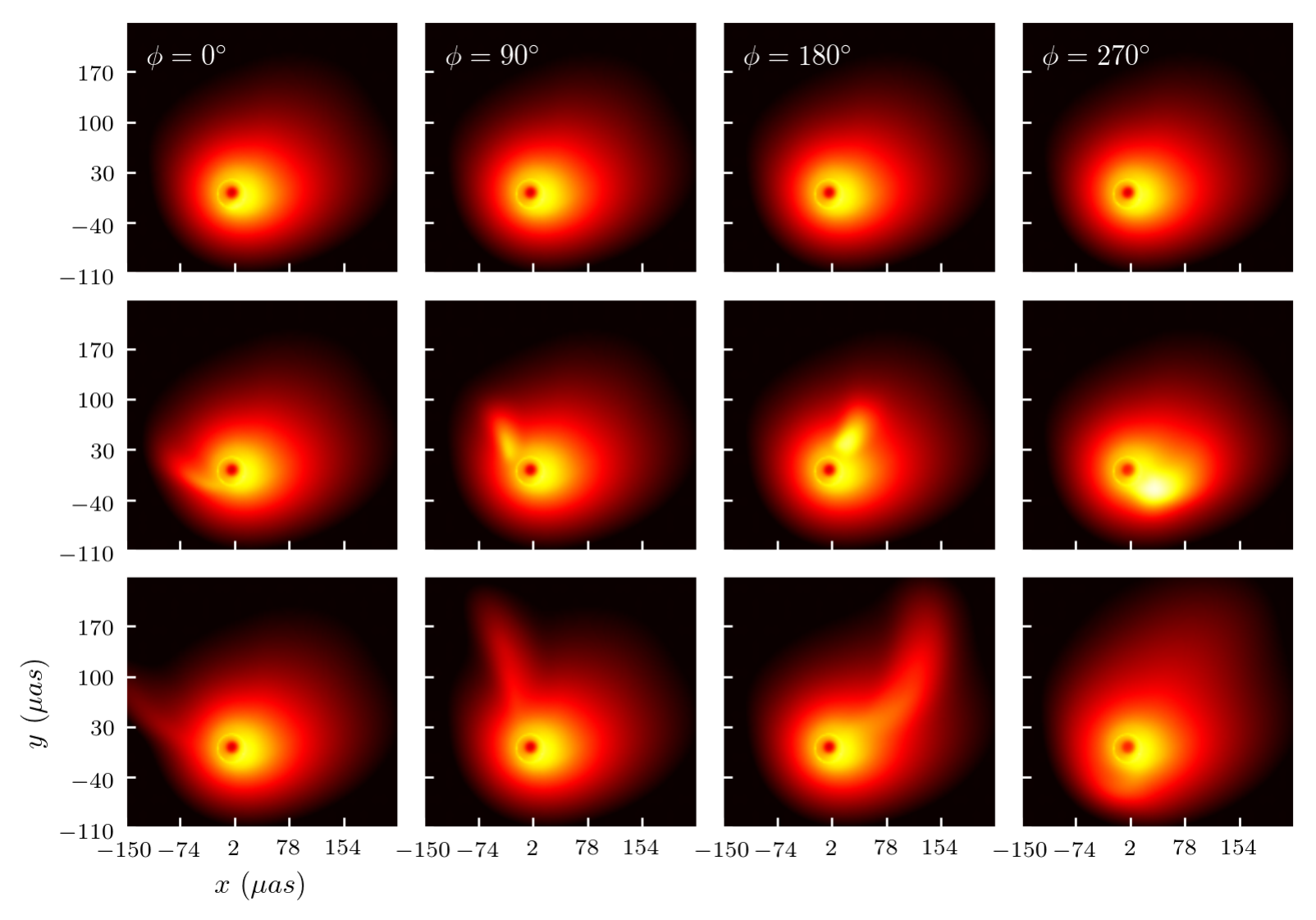} 
  \end{center}
  \caption{Horizon scale images of a spot launched at $\rho=1.5\rho_{\rm crit}$ and at $\phi$ angles of $0^{\circ}$, $90^{\circ}$, $180^{\circ}$, and $270^{\circ}$.  The top row is $5 t_g$ (45 hours) after the spot launch, the second row is $20 t_g$ (7.5 days) after the spot launch, and the bottom row is $35 t_g$ (13 days) after the spot launch.  The spot density remains compact and relatively spherical, and projection effects and light delays produce the dim extended arms at late times.  The spot intensity is generally comparable to the underlying jet intensity.}
    \label{fig:ex150AzZoom}
\end{figure*}

Fixing the radial launch position to $50\%$ the critical surface but varying the azimuthal launch angle around the jet dramatically alters the structure of intensity in the image, especially at late times.  These dramatic differences can be attributed to different projections of the shearing spot as it shears around the jet, and there is qualitatively no structural differences in the local spot density structure between these different launch positions.  The spot in each of these cases shears in the same way and in the same amount of time, as we can see in Figure \ref{fig:ex50AzZoom}.  At late enough times, e.g., the last row of Figure \ref{fig:ex50AzZoom}, there is a thin arc emerging from the edge of the black hole shadow towards the low surface brightness structures further away from the black hole.  This region is the tail of the spot re-entering the region of the jet that provides the strongest beaming.  Even though the spot tail has very low density, the beaming in this region of the jet is strong enough to make the tail of comparable brightness to the main spot arc.  

Similar to black hole-driven spots, launching spots on the critical surface at different azimuthal positions leads to substantially different emission structures in the image at later times, as seen in Figure \ref{fig:ex100AzZoom}.  Even so, these different structures are different projections of the same spot arc around the jet.  Arcs in these images stay relatively compact and bright compared to arcs associated with black hole-driven spots.  Spots on the critical surface shear less than spots inside the critical surface, and no spot on the critical surface shears completely around in the simulated time.  

Also similar to black hole-driven spots, fixing the radial launch position to the critical surface but changing the azimuthal launch position changes the apparent brightness of the spot as it moves around the jet, as we can see in the second row of Figure \ref{fig:ex100AzZoom}.  The spot launched on the side with the highest beaming, $\phi=270^{\circ}$, has a much higher brightness compared to the spot launched at $\phi=0^{\circ}$.  The spot intensity half-life is approximately the same for spots launched on the critical surface as as spots launched inside the jet.   

As mentioned earlier, the consequences of slow light can best be seen in images of wind-driven spots.  We can see in Figure \ref{fig:ex150AzZoom} that for intermediate times, e.g., in the second row, the image of the spot is relatively compact, and approximates the physical structure of the spot at different projections.  At later times, the image of the spot stretches into an extended, diffuse arm even though the spot itself remains relatively spherical.  

Wind-driven spots are generally dimmer relative to interior or critical spots, except on the brightest side of the jet ($\phi=270^{\circ}$).  While these spots may not be as bright as other spots, they contribute significantly to the overall image intensity for much longer.  

\subsection{Critical Spots, Changing Height}

\begin{figure}
 \begin{center}  
  \includegraphics[width=\columnwidth]{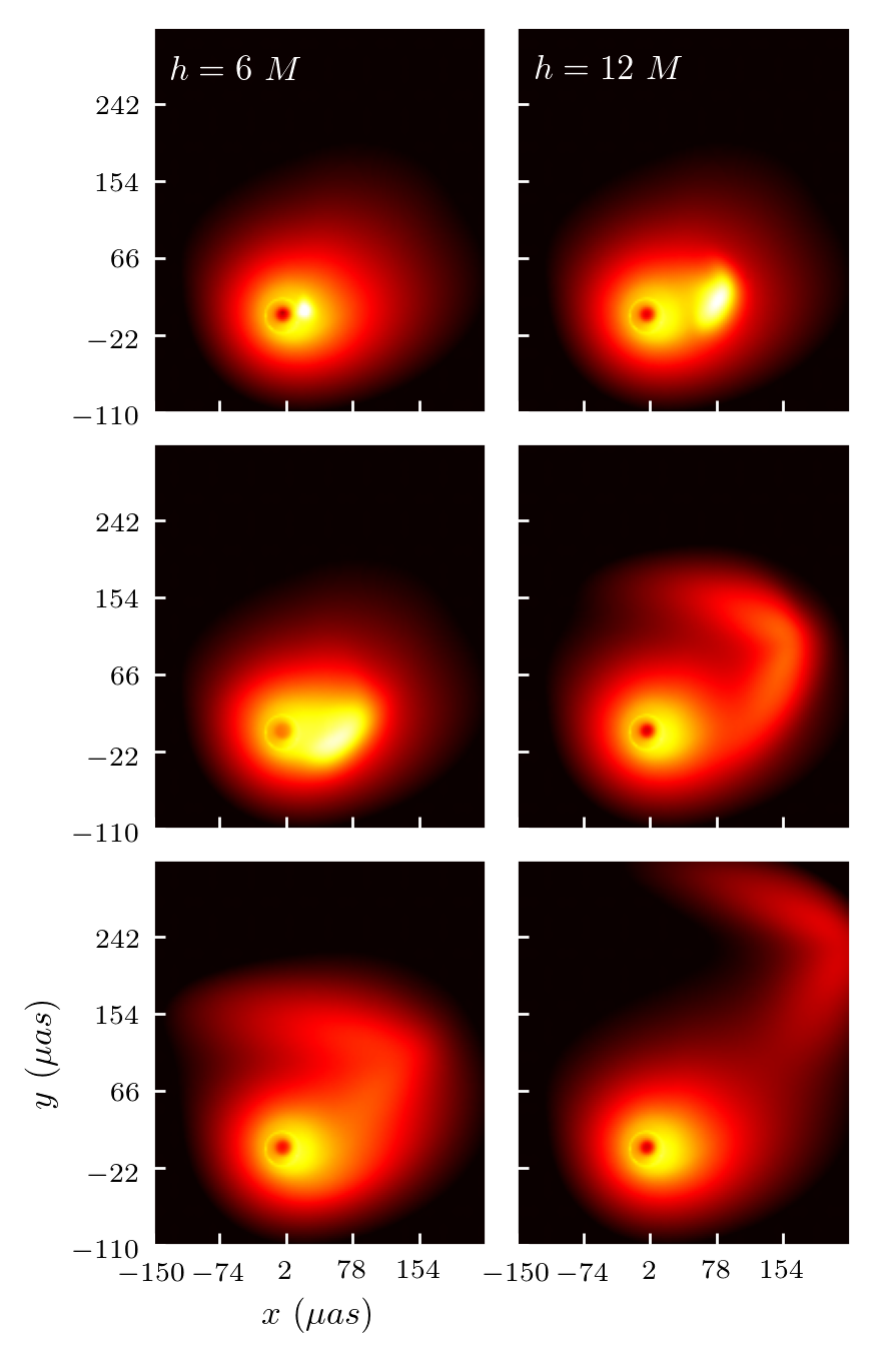} 
 \end{center}
 \caption{ Altering the launch height of spots launched at $\rho=\rho_{\rm crit}$.  The first column is a spot launched with a height of $6 M$ and the second column is a spot launched with a height of $12 M$.  The first row corresponds to $t=5 t_g$ (45 hours) after the spot launch, and the second and third row correspond to $t=15 t_g$ and $t=25 t_g$ respectively.  Early time spots at high launch heights look qualitatively like late time spots with low launch heights.}
 \label{fig:exHeight}
\end{figure}

Fixing the radial and azimuthal launch positions but altering the launch height is approximately degenerate with changing the observer time for the spot image.  Spots that start at a higher position look structurally very similar to later images of spots launched at lower heights.  For example, the spot launched at $12 M$ and seen about 135~hr after launch looks very much like a spot launched at $6 M$ imaged 225~hr after launch (Figure \ref{fig:exHeight}).

\begin{figure}
   \includegraphics[width=\columnwidth]{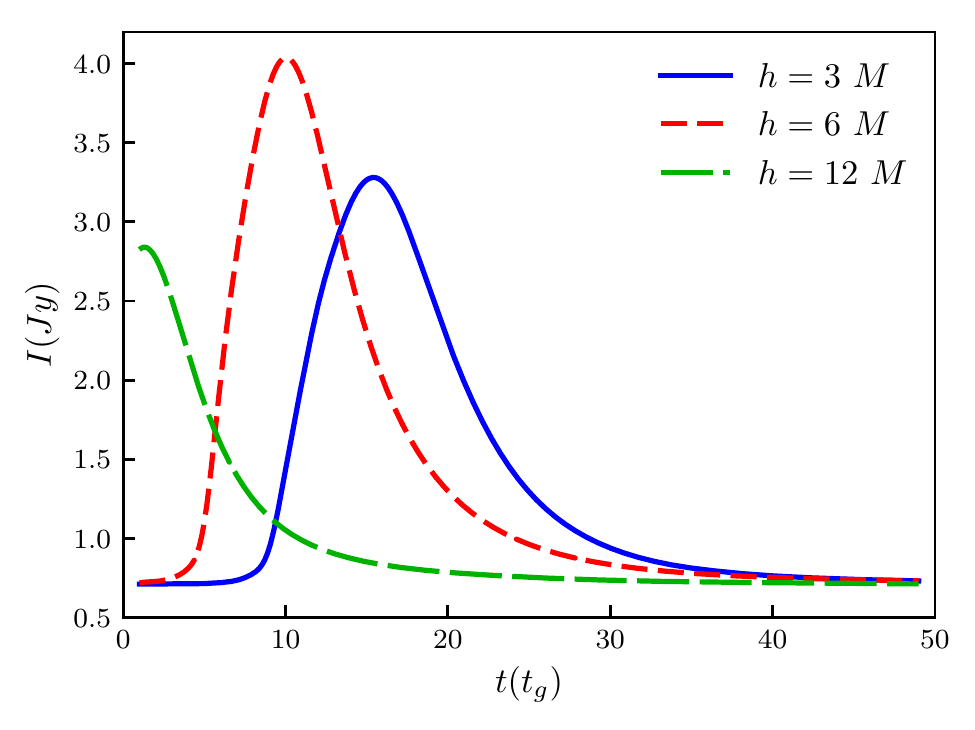}
   \caption{Light curves of a shearing spot launched at $1.0\rho_{\rm crit}$.  The spot was launched at $270^{\circ}$, at heights of $h=3 M$ (blue solid), $h=6 M$ (red dashed), and $h=12 M$ (green long dashed).  Spots that start lower than $6 M$ experience a full rise, peak, and fall in their light curve.}
  \label{fig:exHeightLcurve}
\end{figure}

The time delay signature apparent in the images of spots launched at different heights are also visible in the light curves, as demonstrated in Figure \ref{fig:exHeightLcurve}.  Spots launched at $h=3 M$ start contributing additional flux beyond the quiescent jet around $t=10 t_g$ after the spot launch, and spots launched at $h=6 M$ start brightening only a few $t_g$ after the spot launch.  This delay can be attributed to the shape of the velocity streamlines, which are roughly parabolic.  Close to the black hole, the spot is accelerating up the jet perpendicular to the line of sight, but eventually travels mostly parallel to the line of sight.  Since the intensity is strongly dependent on beaming effects, the spots only contribute to the image intensity when they are traveling parallel to the line of sight.  For spots launched higher up the jet, e.g. at $h=12 M$ ,the spot immediately accelerates parallel to the line of sight.

\subsection{Exploration of Light Curves}

The total averaged image intensity is an easily accessible and measurable quantity that can be accessed without full image reconstructions of the horizon region.  Here we discuss in more detail how the total image intensity changes with spot azimuthal launch position, and how changing the spot launch radius changes the image intensity decay time outside the jet.  

\begin{figure}
  \includegraphics[width=\columnwidth]{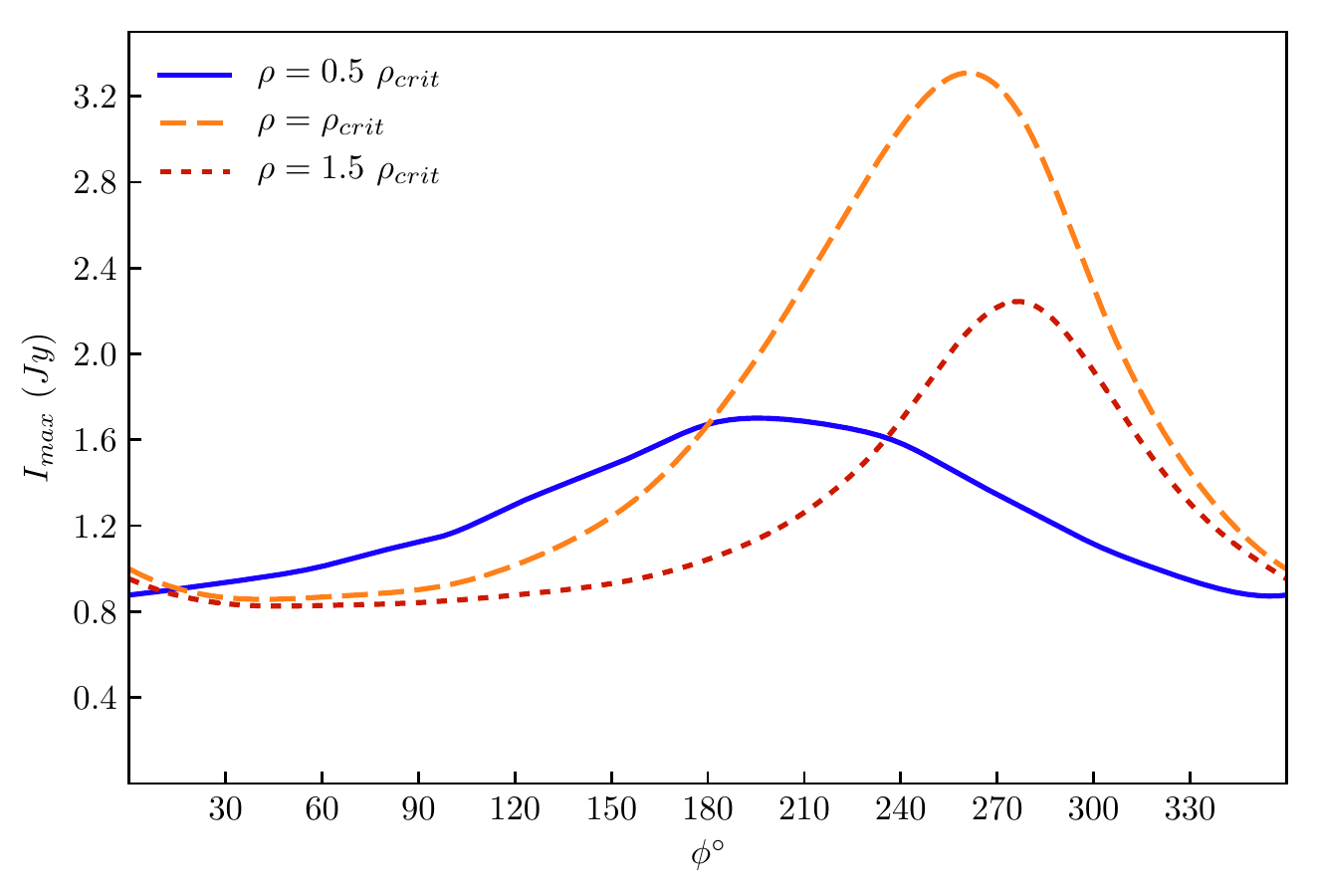} 
  \caption{Maximum image intensity as function of azimuthal launch position $\phi$ for a spot launched at $\rho=0.5\rho_{\rm crit}$ (blue solid line), $\rho_{\rm crit}$ (orange dashed line), and $1.5\rho_{\rm crit}$ (red dotted line).  The maximum intensity peaks near $\phi=270^{\circ}$ for wind driven spots and spots launched on the critical surface, but black hole driven spots have a shallower and broader maximum intensity distribution.}
 \label{fig:MaxIvPhi}
\end{figure}

To explore the effect of different azimuthal launch positions on the structure and maximum intensity in the light curves, we generated a shearing spot simulation every $20^{\circ}$ around the black hole for spots launched at $\rho=0.5,~1.0$, and $1.5\rho_{crit}$.  For each simulation and associated light curve, the maximum image intensity is shown in Figure \ref{fig:MaxIvPhi} as a function of azimuthal launch position, $\phi$.  The quiescent jet without a spot has a total image intensity of about $I=0.7~\Jy$.  The brightest spots are launched between $\phi=210^{\circ}$ and $\phi=300^{\circ}$, and peak around $\phi=270^{\circ}$ for spots launched on the critical surface, as shown by the orange dashed line in Figure \ref{fig:MaxIvPhi}.  Wind driven spots are also brightest around $\phi=270^{\circ}$, but exhibit the lowest maximum intensities compared to other spots when launched between $\phi=10^{\circ}$ and $\phi=240^{\circ}$.  For black hole driven spots launched at $\rho=0.5\rho_{crit}$, the brightest spots are launched closer to $\phi=180^{\circ}$, and only reach about half the peak intensity of spots launched on the critical surface. 

\begin{figure*}
 \begin{center}
  \includegraphics[width=2.1\columnwidth]{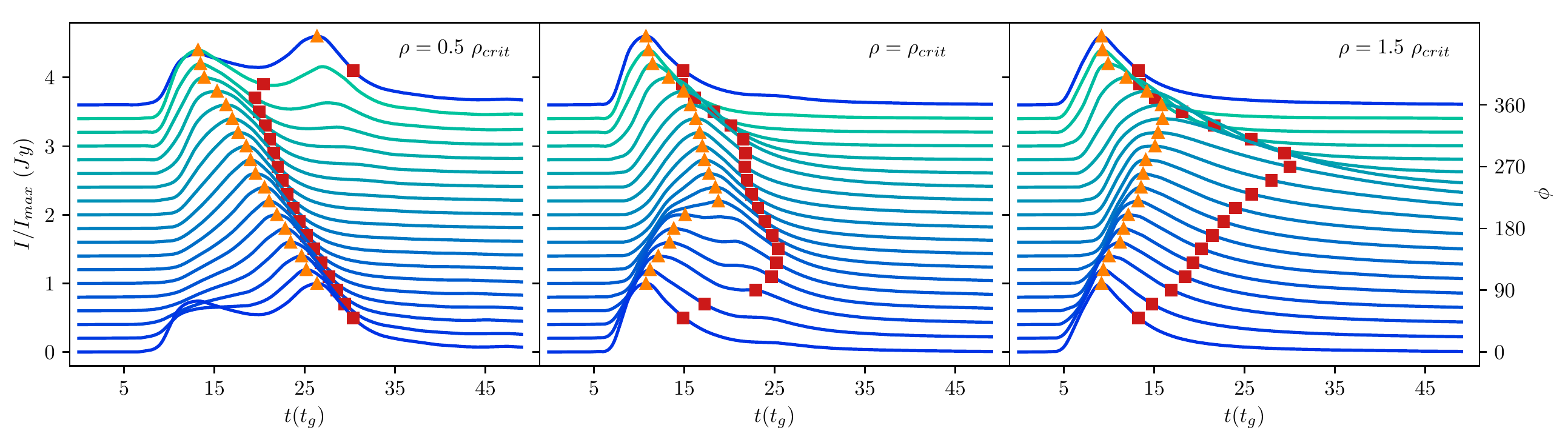} 
  \end{center}
  \caption{Normalized light curves as function of azimuthal launch position $\phi$ for spots launched at $\rho=0.5\rho_{\rm crit}$, $\rho_{\rm crit}$, and $1.5\rho_{\rm crit}$.  Successive light curves are shifted by a constant factor.  The maximum intensity is denoted by an orange triangle, and the half-maximum intensity is denoted by a red square.  Contributions to the light curve from secondary spot features produces non-trivial evolution in the maximum intensity and spot half-life for black hole driven spots.  Light curve features evolve smoothly for wind driven spots.}
  \label{fig:ex100PhiHL}
\end{figure*}

The region of azimuthal parameter space where beaming boosts the image intensity to $>3$ times the quiescent jet occurs between $\phi=200^{\circ}$ and $\phi=300^{\circ}$, about a quarter of the total azimuthal parameter space.  Outside this region, the maximum intensity remains within a factor of 2 or less of the quiescent jet.  Image reconstructions and geometric model fits to M87 presented in \citet{EHT-4:19, EHT-6:19} were fit to visibility amplitudes, which allowed for some measurement of the compact flux in the EHT images.  The average measurement was $0.66^{+0.16}_{-0.10}~\Jy$ for image domain measurements, and $0.75~\pm 0.3 ~\Jy$ for geometric models, but the measurements may suggest a rise in flux during the first two observations and a fall over the last two observations. 

Ideally, to identify and analyze these energetic events, the spot must persist long enough to be seen in at least two observations during the EHT observing window.  The time it takes for a spot to evolve from its maximum intensity to half the maximum is a useful quantity for characterizing the spot lifecycle.  For a spot starting on the critical radius, the spot half-life is largest when the spot starts on the dim side of the jet ($\phi \approx 120^{\circ}$), as see in the centre panel of Figure \ref{fig:ex100PhiHL}.  Here, the spot half-life is approximately 90~hr, but drops to around 45~hr when the spot should be brightest ($\phi=270^{\circ}$).  This variation in spot half-life can be attributed to the presence of secondary features in the light curve caused by the spot tails entering the strongly beamed region.  This appears in all black hole driven spots, and produces breaks and discontinuities in the time of the max and half-maximum intensities.  For a spot launched at $0.5 \rho_{crit}$, this leads to a $10 t_g$ jump between the maximum intensities between spots launched between $\phi=340^{\circ}$ and $\phi=360^{\circ}$.  

Wind driven spots exhibit more gentle light curve evolution with azimuthal launch position.  Both the maximum and half-maximum intensity smoothly rise to peak at $\phi=270^{\circ}$, and then fall back down to the original times at $\phi=0^{\circ}$.  Wind driven spots also exhibit the longest half-lives, where a spot launched at $1.5\rho_{crit}$ has a half-life of about $15 t_g$ when launched at $\phi=240^{\circ}$.  

Within the context of the \citet{EHT-6:19} compact flux measurements, the weak rise and fall over the 6 day observation period does not allow us to put constraints on our spot model at this time.  Most of the uncertainty in the measured flux comes from systematic uncertainty in the calibration of the geometric models to a large library of GRMHD simulations.  If future analyses or observations are able to restrict the allowed GRMHD simulations, and thus contract the systematic error in the flux calibration, then it may be possible to connect the variability at horizon scales to these shearing spot models, and either constrain or rule out certain regions of the spot parameter space.

\begin{figure}
  \includegraphics[width=\columnwidth]{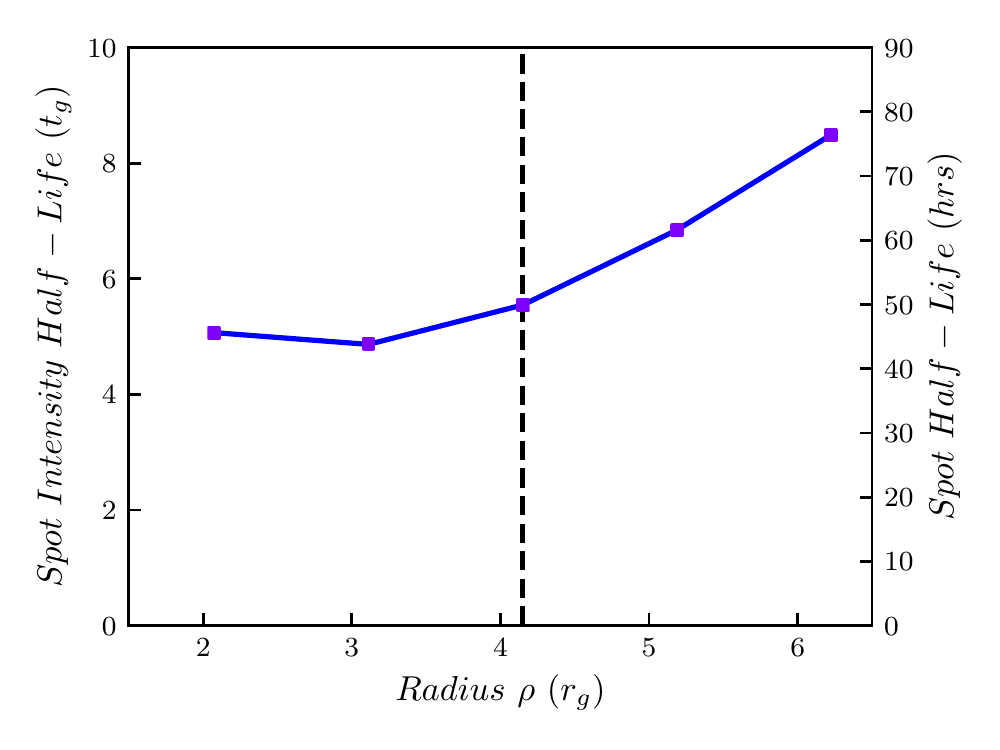} 
  \caption{Spot intensity decay time as function of cylindrical radial launch position $\rho$ for spots launched at $\phi=270^{\circ}$.  The dashed vertical line corresponds to the cylindrical radius of the critical surface.  Spot intensity decay time is relatively flat inside the critical surface but increases outside the critical surface. }
  \label{fig:RadHL}
\end{figure}

At very low azimuthal launch angles ($\phi \approx 0^{\circ}$), all spots have approximately the same maximum intensity and half-life.  However, black hole driven spots may still be distinguishable from wind driven spots by the presence of multi-modal features in the light curves.  When spots are launched at higher azimuthal angles, and especially for spots launched to maximize their intensity, Figure \ref{fig:RadHL} shows that there is a general increasing trend in spot half-lives for increasing cylindrical launch radius.  By using a combination of maximum intensity, spot half-life, and the presence or absence of secondary features in the light curve, it should be possible to distinguish black hole driven and wind driven shearing spots. 

\section{Conclusions}

The emission from compact spots near the base of M87's jet is strongly shaped by the velocity field of the jet.  Material originating inside or on the jet velocity critical surface experience significant shear forces on timescales of a few days, which means consecutive days of EHT observation could observe this type of variability.  These spots are sheared into complex arcs perpendicular to the jet axis that are readily apparent in simulated reconstructed images of the black hole region of M87, and could be verified in the reconstructed EHT images.  Material originating outside the velocity critical surface experience much less shear, producing no arcs in reconstructed images, and are thus distinguishable from interior spots in the image domain.  

The evolution of structure in these spots is strongly dependent on the radial and azimuthal launch position of the material.  While the azimuthal launch position of the spot dramatically alters the maximum intensity of the reconstructed image, the sheared structure remains qualitatively the same for fixed launch radii.  Black hole driven spots exhibit high maximum intensities but fall to quiescent jet intensities within a couple of days, and exhibit multi-modal features in their light curve.  Exterior spots exhibit lower maximum intensities for most azimuthal launch positions but persist for much longer, and can contribute to the light curve for over a week.  Changing the azimuthal launch position of a spot can alter the maximum intensity by up to a factor of 5 due to differences in beaming around the jet.  By combining the spot maximum intensity, half-life, and light curve structure, it is possible to distinguish black hole driven and wind driven spots, and possibly constrain the formation and launching character of AGN jets. 

These distinctions persist in the visibility data, which may be used to probe jet launching physics without using image reconstruction.  In a future publication we hope to explore how visibility data can also help distinguish between black hole driven and wind-driven relativistic jets.

The publication of the black hole image from the EHT observations taken in 2017 signify the start of a new era of horizon-scale science.  While those results primarily focused on image features, direct GRMHD comparisons, and geometric modeling \citep{EHT-4:19, EHT-5:19, EHT-6:19}, incorporating semi-analytic models into the feature-extraction pipeline is a key goal for EHT analyses going forward.  In particular, adding both the force-free jet and shearing spot model as described in this work to the THEMIS model comparison framework (A.E. Broderick et. al 2019, in prep) will allow for precise estimates of jet launching mechanisms and tight constraints on time variability.  The shearing spot model is designed to be compatible in an arbitrary velocity field, and can also be used to model time variability in SMBH systems without jets, or with more exotic velocity profiles.   

\section*{Acknowledgements}
We thank the following for useful discussions: Hung-Yi Pu, Yosuke Mizuno, and Jim Moran.  B.J., A.E.B., and R.G.~receive financial support in part from the Perimeter Institute for Theoretical Physics.   Research at Perimeter Institute is supported by the Government of Canada through Industry Canada and by the Province of Ontario through the Ministry of Research and Innovation.  A.E.B. thanks the Delaney Family for their generous financial support via the Delaney Family John A. Wheeler Chair at Perimeter Institute.  B.J. and A.E.B receive additional financial support from the Natural Sciences and Engineering Research Council of Canada through a Discovery Grant.  R.G. ~also receives support from the ERC synergy grant ``BlackHoleCam: Imaging the Event Horizon of Black Holes'' (Grant No. 610058).

\bibliographystyle{mnras}
\bibliography{m87}
\bsp	
\label{lastpage}
\end{document}